\documentclass[twocolumn,showpacs,preprintnumbers,amsmath,amssymb]{revtex4}

\usepackage{graphicx}
\usepackage{dcolumn}
\usepackage{bm}
\begin{document}

\title{Entanglement swapping between spacelike separated atoms}

\author{Juan Le\'on}
\email{leon@imaff.cfmac.csic.es} \homepage{http://www.imaff.csic.es/pcc/QUINFOG/}

\author{Carlos Sab\'{i}n}%
 \email{csl@imaff.cfmac.csic.es}
\homepage{http://www.imaff.csic.es/pcc/QUINFOG/}
\affiliation{%
Instituto de F\'{i}sica Fundamental, CSIC
 \\
Serrano 113-B, 28006 Madrid, Spain.\\
}%


\date{\today}

\begin{abstract}
We show a mechanism that projects a pair of neutral two-level atoms from an initially uncorrelated state to a maximally
entangled state while they remain spacelike separated. The atoms begin both excited in a common electromagnetic vacuum, and the
radiation is collected with a partial Bell-state analyzer. If the interaction time is short enough and a certain two-photon Bell
state is detected after the interaction, a high degree of entanglement, even maximal, can be generated while one atom is outside
the light cone of the other, for arbitrary large interatomic distances.
\end{abstract}

\maketitle


\section{Introduction}
Entanglement between distant atoms is a key resource for Quantum Information and Computation. There are mainly two different
known ways of generate it: by interaction between the atoms (for instance, \cite{vanenk}) or by  detection of the emitted
photons \cite{cabrillo,feng,duan,simon,lamata}. Some of these proposals have been realized experimentally (for instance,
\cite{moehring}). For the latter cases, in principle, there is no reason to expect that the swapping \cite{swapping} of
atom-photon to atom-atom entanglement can only begin to occur when one atom enter into the light cone of the other.

The possibility of entanglement generation between spacelike separated atoms is of both theoretical and practical interest, and
was addressed from different points of view in \cite{franson,reznik,reznikII,conjuan}. In \cite{conjuan}, we analyze this issue
perturbatively in a simple model of a pair of two-level atoms interacting locally with the electromagnetic field, initially in
the vacuum state \cite{fermi}. Tracing over the field states, the atoms are only classically correlated, but applying
$|\,n\rangle\langle\,n|$ ($n\,=\,0, 1, 2$ being the number of photons up to second order in perturbation theory), the atoms get
entangled. For $n\,=\,0$ the entanglement is generated by the interaction term and therefore is only relevant when one atom
enter into the light cone of the other, despite of the finiteness of the Feynman propagator beyond that region. But for $n\,=\,
1, 2$ entanglement may be sizeable, although small, if the interatomic distance is short enough. In \cite{reznik}, the trace
over the field states was considered in a model with a pair of two-level detectors coupled to a scalar field. The detectors may
get entangled if a suitable time dependent coupling is introduced, and this was applied to a linear ion trap in \cite{reznikII}.
In \cite{franson}, only the vacuum case when $t\rightarrow0$ was analyzed, and no entanglement measures were considered.

There are two possible interpretations for these effects: as a transfer of preexisting entanglement of the vacuum \cite{reznik,
reznikII} or as a consequence of the propagation of virtual quanta outside the light cone \cite{franson}. Both are compared and
discussed in \cite{franson}.

In this paper we will go one step further and consider that the photons are detected with definite momenta and polarizations. We
show that, in principle, a high degree of entanglement, even maximal, can be generated between spacelike separated atoms if a
Bell state of the emitted photons is detected. We will consider a pair of neutral two-level atoms separated by a fixed and
arbitrary distance and study the evolution of an initially uncorrelated state under local interaction with the electromagnetic
field. We focus on the two-photon emission which, although has a smaller probability of success, shows a larger fidelity of the
projected state with the desired state and has a entanglement robust to atomic recoil \cite{morigi}. The photons pass through a
partial Bell-state analyzer \cite{zeilinger}, and we use entanglement measures to study the evolution of entanglement in the
projected atomic states after detection of the different photonic Bell states. The results show that interaction times must be
short, but interatomic distances can be as large as desired. The interaction time is independent of the photodetection time,
which is only related with the distance from the atoms to the detectors.  That distance can be such that the photodetection can
occur while the atoms remain spacelike separated.

The results can be interpreted as a transfer of part of the vacuum entanglement after a post-selection process. If no
measurement were performed the atoms would have classical correlations transferred by the vacuum. In \cite{reznik} the classical
correlations may become entanglement with a suitable time dependent coupling. The post-selection process can be seen as an
alternative way to achieve the entanglement transference. While the results in \cite{franson,reznik} are mainly theoretical,
these could be probed experimentally, and would show for the first time the possibility of transfer entanglement from the vacuum
state of the quantum field to spacelike separated atoms.

\section{Entanglement swapping between spacelike separated atoms}

To address the atom-field interactions, we assume that the relevant wavelengths and the interatomic separation are much larger
than the atomic dimensions.  The dipole approximation, appropriate to these conditions, permits the splitting of the system
Hamiltonian into two parts $H = H_0 + H_I$ that are separately gauge invariant. The first part is the Hamiltonian in the absence
of interactions other than the potentials that keep $A$ and $B$ stable and the selfinteraction terms that can be removed when
radiative corrections are considered \cite{cohentannoudji}, $H_0 = H_A + H_B + H_{\mbox{field}}$. The second contains all the
interaction of the atoms with the field $H_I = - \frac{1}{\epsilon_0}\sum_{n=A,B}
\mathbf{d}_n(\mathbf{x}_n,t)\,\mathbf{D}(\mathbf{x}_n,t)$,where $\mathbf{D}$ is the electric displacement field, and
$\mathbf{d}_n \,=\,\sum_i\, e\,\int d^3 \mathbf{x}_i\, \langle\,E\,|\,(\mathbf{x}_i-\mathbf{x}_n)\,|\,G\,\rangle$ is the
electric dipole moment of atom $n$, that we will take as real and of equal magnitude for both atoms
$(\mathbf{d}=\mathbf{d_A}=\mathbf{d_B})$, $|\,E\,\rangle$ and $|\,G\,\rangle$ being the excited and ground states of the atoms,
respectively.

In what follows we choose a system  given initially by the product state, $|\,\psi\,\rangle_0\,=\,
|\,E\,E\,\rangle\cdot|\,0\,\rangle$ in which atoms $A$ and $B$ are in the excited state $|\,E\,\rangle$  and the field in the
vacuum state $|\,0\,\rangle$. The system then evolves under the effect of the interaction during a lapse of time $t$, and, up to
order $e^2$, 0, 1 or 2 photons may be emitted. If after that a two-photon state is detected,
\begin{equation}
|\Psi\rangle=|\mbox{photon}_1,\mbox{photon}_2\rangle=\sum_{\vec{k},\vec{k'},\lambda,\lambda'}\,c_{\vec{k}\,\vec{k'},\lambda,\lambda'}\,|\vec{k}\lambda,\,\vec{k'}\lambda'\rangle
\label{a}
\end{equation}
(being $\hbar\,\vec{k}$, $\hbar\,\vec{k'}$ momenta and $\lambda$,$\lambda'$ polarizations), the projected state, up to order
$e^2$, can be given in the interaction picture as
\begin{equation}
|\mbox{photons},\mbox{atom}_1,\mbox{atom}_2\rangle_{t} = |\,\Psi\rangle(\frac{f\,|\,E\,E\rangle+ g\,|\,G\,G\rangle}{N})
\label{b}
\end{equation}
where
\begin{equation}
f=+\frac{1}{2}\langle\Psi|T(\mathcal{S}_A^+ \mathcal{S}_A^-\,+\, \mathcal{S}_B^+\mathcal{S}_B^-)|0\rangle,\,
g=+\langle\Psi|T(\mathcal{S}^-_B \mathcal{S}^-_A)|0\rangle \label{c}
\end{equation}
and $N=\sqrt{|\,f\,|^2+|\,g\,|^2}$, being $\mathcal{S}\,=\,- \frac{1}{\hbar}  \int_0^t\, dt'\, H_{I}(t')$
$(\mathcal{S}=\mathcal{S}^{+}\, +\, \mathcal{S}^{-})$, and $T$ the time ordering operator. Here, $g$  describes single photon
emission by both atoms, while $f$ corresponds to two photon emission by a single atom. The sign of the superscript is associated
to the energy difference between the initial and final atomic states of each emission. In Quantum Optics, $f$ is usually
neglected by the introduction of a rotating wave approximation (RWA), but as we will see later, for very short interaction times
$f$ and $g$ may be of similar magnitude. Actually, a proper analysis of this model can be performed only beyond the RWA
\cite{powerthiru,milonni,compagnoI}. Without RWA vacuum entanglement cannot be transferred to the atoms with this particular
post-selection process, but the trace over the field states would be a classically correlated state, as in \cite{conjuan}. In
that case, a one photon post-selection process would entangle the atoms.

(\ref{c}) can be written as:
\begin{eqnarray}
f&=&+\frac{1}{2}\,\theta(t_1-t_2)\langle\Psi|\mathcal{S}_A^+\,(t_1)
\mathcal{S}_A^-\,(t_2)+\mathcal{S}_B^+\,(t_1)\mathcal{S}_B^-\,(t_2))|0\rangle,\,\nonumber\\
g&=&+\langle\Psi|\mathcal{S}^-_B\,(t_1) \mathcal{S}^-_A\,(t_2))|0\rangle \label{d}
\end{eqnarray}

Finally, in the dipole approximation the actions $\hbar\, \mathcal{S}^{\pm}$ in (\ref{d}) reduce to
\begin{eqnarray}
\mathcal{S}^{\pm}\,=\,- \frac{i}{\hbar}  \int_0^t\, dt' \: e^{\pm i\Omega t'}\,
\mathbf{d}\cdot\mathbf{E}(\mathbf{x},t')\label{e}
\end{eqnarray}
where  $\Omega = \omega_E -\omega_G$ is the transition frequency, and we are neglecting atomic recoil. This depends on the
atomic properties $\Omega$ and $\mathbf{d}$, and on the interaction time $t$. In our calculations we will take $(\Omega
|\mathbf{d}|/e c) = 5\,\cdot 10^{-3}$, which is of the same order as the 1s $\rightarrow$ 2p transition in the hydrogen atom,
consider $\Omega t \gtrsim 1$, and analyze the cases $(L/c\,t)\simeq 1$ around the light cone, $L$ being the interatomic
distance. We will use the standard mode expansion for the electric field: $\mathbf{E}(\mathbf{x})=i\sqrt{\frac{\hbar\,
c}{2\varepsilon_0\,(2\pi)^3}}\sum_{\lambda}\int d^3k
\sqrt{k}(e^{i\mathbf{k}\,\mathbf{x}}\mathbf{\epsilon}(\mathbf{k},\lambda)\,a_{k\lambda}-e^{-i\mathbf{k}\,\mathbf{x}}\mathbf{\epsilon}^*\,(
\mathbf{k},\lambda)\,a^{\dag}_{k\lambda})$, with
$[\,a_{k\lambda},\,a^{\dag}_{k'\lambda'}\,]=\delta^3(\mathbf{k}-\mathbf{k'})\,\delta_{\lambda\,\lambda'}$.

The photons pass through a partial Bell-state analyzer \cite{zeilinger} consisting in a beam splitter (BS) and two polarization
beam splitters (PBS) with four single photon detectors at their output ports. If two detectors, one at one output port of one
PBS and one at an output port of the other, click at the same time, a state $|\Psi^-\rangle$ is detected, while if the two
clicks are in the two output ports of only one PBS, the state is $|\Psi^+\rangle$. If one of the four detectors emits a double
click, the state can be $|\Phi^+\rangle$ or $|\Phi^-\rangle$. Taking into account momenta and symmetrization, the Bell states
can be written as
\begin{eqnarray}
|\Psi^{\pm}\rangle&=&\frac{1}{\sqrt2}[|\vec{k}\downarrow,\,\vec{k'}\uparrow\rangle+|\vec{k'}\uparrow,\,\vec{k}\downarrow\rangle\nonumber\\
&\pm&(|\vec{k}\uparrow,\,\vec{k'}\downarrow\rangle+|\vec{k'}\downarrow,\,\vec{k}\uparrow\rangle)]\label{f}\\
|\Phi^{\pm}\rangle&=&\frac{1}{\sqrt2}[|\vec{k}\downarrow,\,\vec{k'}\downarrow\rangle+|\vec{k'}\downarrow,\,\vec{k}\downarrow\rangle\nonumber\\
&\pm&(|\vec{k}\uparrow,\,\vec{k'}\uparrow\rangle +|\vec{k'}\uparrow,\,\vec{k}\uparrow\rangle)]\nonumber
\end{eqnarray}
where $\uparrow$ and $\downarrow$  are the photon polarizations, with polarization vectors
$\epsilon\,(\vec{k},\uparrow)={-1\over\sqrt2}(\epsilon\,(\vec{k},1)+\epsilon\,(\vec{k},2))$ and
$\epsilon\,(\vec{k},\downarrow)=$ ${1\over\sqrt2}(\epsilon\,(\vec{k},1)-\epsilon\,(\vec{k},2))$, where $\epsilon\,(\vec{k},1)=$
$(\cos{\theta_k}\cos{\phi_k},\cos{\theta_k}\sin{\phi_k},-\sin{\theta_k})$ and $\epsilon\,(\vec{k},2)=$
$(-\sin{\phi_k},\cos{\phi_k},0)$. Here $|\vec{k}\lambda,\,\vec{k'}\lambda'\rangle=a^{\dag}_{k\lambda}\,
\,a^{\dag}_{k'\lambda'}|\,0\rangle$.

We will use the concurrence \cite{wootters} $\mathbb{C}$ to compute the entanglement of the atomic states when the different
Bell states are detected. The concurrence of the atomic part of a state like (\ref{b}) is just given by
\begin{equation}
\mathbb{C}=\frac{2 |\,f\,g^*\,|}{N^2} \label{g}
\end{equation}
We assume that the atoms $A$, $B$ are along the $y$ axis, at $y=\,\mp L/2$ respectively, and the dipoles are parallel along the
$z$ axis, corresponding to an experimental set up in which the dipoles are induced by suitable external fields \cite{franson}.
We also take $|\vec{k}|=|\vec{k'}|=\Omega/c$.

Under that conditions, the first remarkable thing is that for $|\Psi^-\rangle$ and $|\Phi^-\rangle$,  we have $f=\,g=\,0$.
Therefore, at least while only E1 transitions are considered, in this model the Bell-sate analyzer is complete: if two different
detectors click the state is $|\Psi^+\rangle$, while if one detector clicks twice the state is $|\Phi^+\rangle$. First, we focus
on $|\Psi^+\rangle$.   Considering (\ref{d}), (\ref{e}) and (\ref{f}), with the mode expansion for the electric field and the
commutation relation for the creation and annihilation operators, a standard computation leads to:
\begin{eqnarray}
f&=&\frac{K(\Omega,t,\theta)}{2}\,j(\Omega\,t)\cos{(\frac{z}{2}\,h_+(\theta,\phi))}\nonumber\\
g&=& K(\Omega,t,\theta)\,\cos{(\frac{z}{2}\,h_-(\theta,\phi))},\label{h}
\end{eqnarray}
with $K(\Omega,t,\theta)=$ $\frac{c\,\alpha|\mathbf{d}|^2\Omega t^2}{2\pi^2e^2}\sin{\theta_k}\,\sin{\theta_{k'}}$, ($\alpha$
being the fine structure constant), $j(\Omega\,t)=$ ${|-1\,+\,e^{2i\,\Omega\,t}(1-2i\,\Omega\,t)|\over(\Omega\,t)^2},$
$h_\pm\,(\theta,\,\phi)=$ $(\sin{\theta_k}\sin{\phi_k}\pm\sin{\theta_{k'}}\sin{\phi_{k'}}),$ $\theta_k,\,\phi_k$ corresponding
to $\hat{k}$ and $\theta_{k'},\,\phi_{k'}$ to $\hat{k'}$, and $z=\,\Omega\,L/c$. Notice that $j(\Omega\,t)$ decreases as $t$
grows, and eventually vanish as $t\rightarrow\infty$ as required by energy conservation. The $L$ dependence is a result of the
individual dependence on the position of each atom, not on the relative distance between them.

Taking into account (\ref{g}) and (\ref{h}) the concurrence is given by:
\begin{equation}
\mathbb{C}={4|\cos{(\frac{z}{2}\,h_+(\theta,\,\phi))}\cos{(\frac{z}{2}\,h_-(\theta,\,\phi))}|\over\cos^2{(\frac{z}{2}\,h_+(\theta,\,\phi))}\,j(\Omega\,t)
+\cos^2{(\frac{z}{2}\,h_-(\theta,\,\phi))}\,{4\over j(\Omega\,t)}}, \label{i}
\end{equation}

Now, we assume that the  50:50 BS is at $(y,z)=(0,L/2)$, the two PBS at $(\pm\,d/2\sqrt2,L/2+d/2\sqrt2)$ and the four detectors
at $(\pm\,d/\sqrt2,L/2+d/\sqrt2)$ and $(\pm\,d/\sqrt2,L/2)$ (see Fig.1). (\ref{h}) will not depend on the value of $d$, which is
the distance traveled by the photon to any detector after leaving the BS. Notice that, with this setup, $h_-=0$ and
$h_+=\sqrt2$.
\begin{figure}[h]
\includegraphics[width=0.75\textwidth]{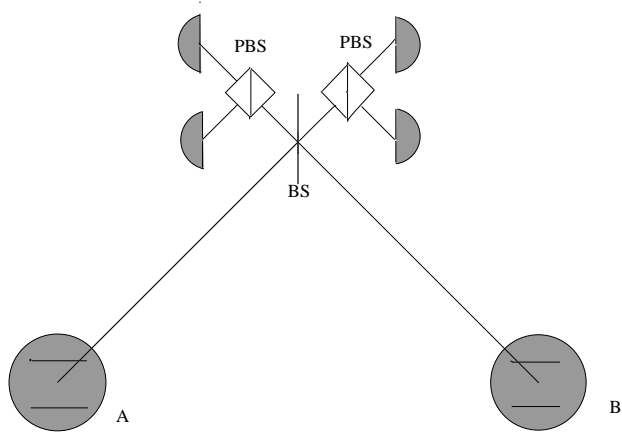}
\caption{Schematic setup for the entanglement swapping described in the text. The atoms $A$ and $B$  are at $(y,z)=(\mp L/2,0)$.
The emitted photons pass through a $50:50$ BS at $(0,L/2)$ and two PBS at $(\pm\,d/2\sqrt2,L/2+d/2\sqrt2)$, and there are four
single photon detectors at the outport ports of the two PBS, at $(\pm\,d/\sqrt2,L/2+d/\sqrt2)$ and $(\pm\,d/\sqrt2,L/2)$. Taking
into account that $|\Psi^-\rangle$ and $|\Phi^-\rangle$ are forbidden in our model, a $|\Psi^+\rangle$ is detected when there
are coincidence clicks in two detectors and $|\Phi^+\rangle$ when there is a double click in one detector. Then the atoms are
projected into the atomic part of the state (\ref{b}).}
\end{figure}

In Fig. 2 we represent (\ref{i}) under that conditions as a function of $x=L/c\,t$ for three different values of $z$ (different
values of $L$). Notice that a high degree of entanglement, maximal for $x$ large enough (short enough interaction times $t$),
can be achieved in all cases when one atom is beyond the light cone of the other ($x>1$).
\begin{figure}[h]
\includegraphics[width=0.45\textwidth]{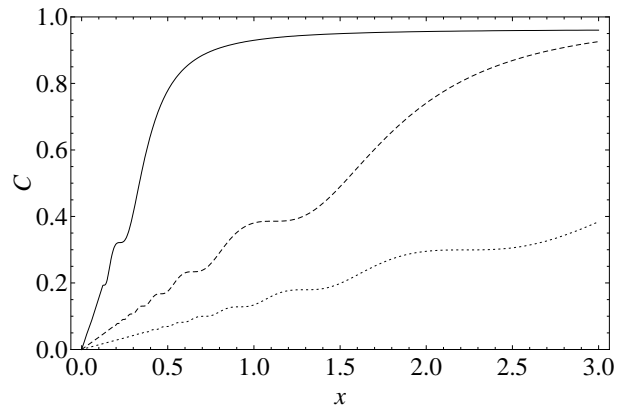}
\caption{Concurrence for the atomic state when a Bell state $|\Psi^+\rangle$ or $|\Phi^+\rangle$ of the photons is detected, as
a function of $x={L/ct}$ for $z=\Omega L/c=$1 (solid), 5 (dashed), 10 (dotted). The light cone is at $x<1$. For $x>1$ the
interaction time is short enough to have a significative amount of entanglement.}
\end{figure}
As $t\rightarrow\infty$ $(x\rightarrow0)$, the concurrence eventually vanish, in agreement with the fact that the only atomic
state allowed by energy conservation is just the separable state $|\,G\,G\,\rangle$.
\begin{figure}[h]
\includegraphics[width=0.45\textwidth]{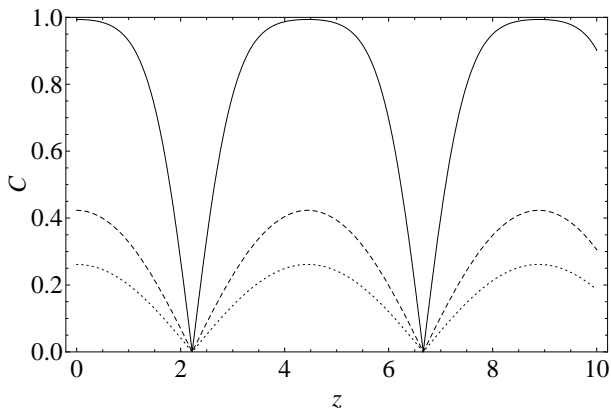}
\caption{Concurrence for the atomic state when a Bell state $|\Psi^+\rangle$  of the photons is detected, as a function of
$z=\Omega L/c$ for $\Omega\,t=$1 (solid), 4 (dashed), 7 (dotted). The light cone for each curve is at $z<\Omega\,t$ .}
\end{figure}

In Fig. 3 we represent (\ref{i}) as a function of $z$ for three different values of $z/x=\Omega\,t$, to give an alternative
description. The mutual light cone corresponds to the region $z<\Omega\,t$ in each case. The concurrence oscillates with the
position of the atoms, and eventually vanish at $z=\sqrt2\,(n+1/2)\,\pi$ ($n=0,1,2...$), as a consequence of the vanishing of
$\cos{(z/\sqrt2)}$. For a given interaction time $t$, the maximum of the concurrence can be achieved for interatomic distances
as large as desired. In particular, a maximally entangled state is generated for $\Omega\, t=1$, which corresponds to
$t\simeq10^{-15}\, s$.

In Fig. 4 we sketch (\ref{i}) as a function of $\phi=\phi_k=\phi_{k'}$ for given values of $x$ and $z$. Notice that the maximum
values for the entanglement are around $\phi=n\,\pi/2$ ($n=0,1,2...$), $\pi/2$ corresponding to the setup of Fig. 1.
\begin{figure}[h]
\includegraphics[width=0.45\textwidth]{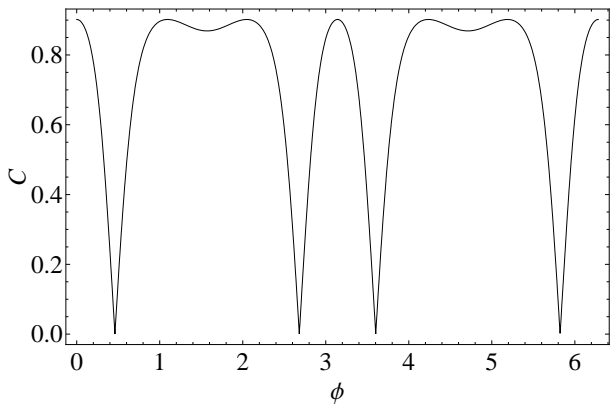}
\caption{Concurrence for the atomic state when a Bell state $|\Psi^+\rangle$ of the photons is detected, as a function of $\phi$
for $z=\Omega L/c=5$ and $x=L\,c/t=2.5$.}
\end{figure}

So far, we have focused on $|\Psi_+\rangle$, but, in principle, $|\Phi_+\rangle$ could be detected as well. The coefficients $f$
and $g$ would have opposite sign to those of $|\Psi_+\rangle$ and therefore the concurrence would be the same. But, due to the
interaction times considered here, the relaxation time of a single detector must be extremely short in order to emit a double
click.

\section{Conclusions}

In conclusion, we have shown that, in principle, two neutral two-level atoms can evolve from an initially uncorrelated state to
a highly entangled state in a time shorter than the time required for the light to travel between them. At the initial time,
both atoms are excited in a common electromagnetic vacuum. They are allowed to interact with the field due to an induced dipole
during a time $t$  and, up to second order in perturbation theory, $n=0,1,2$ photons may be emitted. After that, the emitted
radiation pass through a partial Bell-state analyzer. For interaction times $t\simeq10^{-15} s$ and if a two-photon Bell state
$|\Psi^+\rangle$ or $|\Phi^+\rangle$ (the other two are forbidden in this model) is detected after that, the atoms are projected
into an entangled state, which may be maximally entangled for short enough $t$. For a given $t$, the degree of entanglement
oscillates periodically with the distance and the maximum degree available can be achieved for interatomic distances $L$ as
large as desired. Notice that the interaction time $t$, which must be $t\simeq10^{-15}\, s$, is absolutely independent of the
time $t'$ at which the photodetection takes place. Since the distance traveled by the photons from the atoms to the detector is
$L/\sqrt2 + d$, $d$ being arbitrary, the photodetection can occur after a time $t'\lesseqgtr L/c$. A suitable choice of $d$ is
necessary in order to ensure that the atoms may remain spacelike separated. The degree of entanglement is independent of $d$.

\begin{acknowledgments}

This work was supported by Spanish MEC FIS2005-05304 and CSIC 2004 5 OE 271 projects.
\end{acknowledgments}


\begin{thebibliography}{20}
\expandafter\ifx\csname natexlab\endcsname\relax\def\natexlab#1{#1}\fi \expandafter\ifx\csname bibnamefont\endcsname\relax
  \def\bibnamefont#1{#1}\fi
\expandafter\ifx\csname bibfnamefont\endcsname\relax
  \def\bibfnamefont#1{#1}\fi
\expandafter\ifx\csname citenamefont\endcsname\relax
  \def\citenamefont#1{#1}\fi
\expandafter\ifx\csname url\endcsname\relax
  \def\url#1{\texttt{#1}}\fi
\expandafter\ifx\csname urlprefix\endcsname\relax\def\urlprefix{URL }\fi \providecommand{\bibinfo}[2]{#2}
\providecommand{\eprint}[2][]{\url{#2}}

\bibitem[{\citenamefont{van Enk et~al.}(1997)\citenamefont{van Enk, Cirac, and
  Zoller}}]{vanenk}
\bibinfo{author}{\bibfnamefont{S.~J.} \bibnamefont{van Enk}},
  \bibinfo{author}{\bibfnamefont{J.~I.} \bibnamefont{Cirac}}, \bibnamefont{and}
  \bibinfo{author}{\bibfnamefont{P.}~\bibnamefont{Zoller}},
  \bibinfo{journal}{Phys. Rev. Lett.} \textbf{\bibinfo{volume}{78}},
  \bibinfo{pages}{4293} (\bibinfo{year}{1997}).

\bibitem[{\citenamefont{Cabrillo et~al.}(1999)\citenamefont{Cabrillo, Cirac,
  P.Garc{\'i}a-Fern{\'a}ndez, and Zoller}}]{cabrillo}
\bibinfo{author}{\bibfnamefont{C.}~\bibnamefont{Cabrillo}},
  \bibinfo{author}{\bibfnamefont{J.~I.} \bibnamefont{Cirac}},
  \bibinfo{author}{\bibnamefont{P.Garc{\'i}a-Fern{\'a}ndez}}, \bibnamefont{and}
  \bibinfo{author}{\bibfnamefont{P.}~\bibnamefont{Zoller}},
  \bibinfo{journal}{Phys.\ Rev.\ A} \textbf{\bibinfo{volume}{59}},
  \bibinfo{pages}{001025} (\bibinfo{year}{1999}).

\bibitem[{\citenamefont{Feng et~al.}(2003)\citenamefont{Feng, Zhang, Li, Gong,
  and Xu}}]{feng}
\bibinfo{author}{\bibfnamefont{X.~L.} \bibnamefont{Feng}},
  \bibinfo{author}{\bibfnamefont{Z.~M.} \bibnamefont{Zhang}},
  \bibinfo{author}{\bibfnamefont{X.~D.} \bibnamefont{Li}},
  \bibinfo{author}{\bibfnamefont{S.~Q.} \bibnamefont{Gong}}, \bibnamefont{and}
  \bibinfo{author}{\bibfnamefont{Z.~Z.} \bibnamefont{Xu}},
  \bibinfo{journal}{Phys. Rev. Lett.} \textbf{\bibinfo{volume}{90}},
  \bibinfo{pages}{217902} (\bibinfo{year}{2003}).

\bibitem[{\citenamefont{Duan and Kimble}(2003)}]{duan}
\bibinfo{author}{\bibfnamefont{L.~M.} \bibnamefont{Duan}} \bibnamefont{and}
  \bibinfo{author}{\bibfnamefont{H.~J.} \bibnamefont{Kimble}},
  \bibinfo{journal}{Phys. Rev. Lett.} \textbf{\bibinfo{volume}{90}},
  \bibinfo{pages}{253601} (\bibinfo{year}{2003}).

\bibitem[{\citenamefont{Simon and Irvine}(2003)}]{simon}
\bibinfo{author}{\bibfnamefont{C.}~\bibnamefont{Simon}} \bibnamefont{and}
  \bibinfo{author}{\bibfnamefont{W.~T.~M.} \bibnamefont{Irvine}},
  \bibinfo{journal}{Phys. Rev. Lett.} \textbf{\bibinfo{volume}{91}},
  \bibinfo{pages}{110405} (\bibinfo{year}{2003}).

\bibitem[{\citenamefont{Lamata et~al.}(2007)\citenamefont{Lamata,
  Garc{\'i}a-Ripoll, and Cirac}}]{lamata}
\bibinfo{author}{\bibfnamefont{L.}~\bibnamefont{Lamata}},
  \bibinfo{author}{\bibfnamefont{J.~J.} \bibnamefont{Garc{\'i}a-Ripoll}},
  \bibnamefont{and} \bibinfo{author}{\bibfnamefont{J.~I.} \bibnamefont{Cirac}},
  \bibinfo{journal}{Phys.\ Rev.\ Lett.} \textbf{\bibinfo{volume}{98}},
  \bibinfo{pages}{010502} (\bibinfo{year}{2007}).

\bibitem[{\citenamefont{Moehring et~al.}(2007)\citenamefont{Moehring, Maunz,
  Olmschenk, Younge, Matsukevich, Duan, and Monroe}}]{moehring}
\bibinfo{author}{\bibfnamefont{D.~L.} \bibnamefont{Moehring}},
  \bibinfo{author}{\bibfnamefont{P.}~\bibnamefont{Maunz}},
  \bibinfo{author}{\bibfnamefont{S.}~\bibnamefont{Olmschenk}},
  \bibinfo{author}{\bibfnamefont{K.~C.} \bibnamefont{Younge}},
  \bibinfo{author}{\bibfnamefont{D.~N.} \bibnamefont{Matsukevich}},
  \bibinfo{author}{\bibfnamefont{L.~M.} \bibnamefont{Duan}}, \bibnamefont{and}
  \bibinfo{author}{\bibfnamefont{C.}~\bibnamefont{Monroe}},
  \bibinfo{journal}{Nature} \textbf{\bibinfo{volume}{68}}, \bibinfo{pages}{449}
  (\bibinfo{year}{2007}).

\bibitem[{\citenamefont{{\.{Z}}ukowski
  et~al.}(1993)\citenamefont{{\.{Z}}ukowski, Zeilinger, Horne, and
  Ekert}}]{swapping}
\bibinfo{author}{\bibfnamefont{M.}~\bibnamefont{{\.{Z}}ukowski}},
  \bibinfo{author}{\bibfnamefont{A.}~\bibnamefont{Zeilinger}},
  \bibinfo{author}{\bibfnamefont{M.~A.} \bibnamefont{Horne}}, \bibnamefont{and}
  \bibinfo{author}{\bibfnamefont{A.~K.} \bibnamefont{Ekert}},
  \bibinfo{journal}{Phys.\ Rev.\ Lett.} \textbf{\bibinfo{volume}{71}},
  \bibinfo{pages}{4287} (\bibinfo{year}{1993}).

\bibitem[{\citenamefont{Franson}(2008)}]{franson}
\bibinfo{author}{\bibfnamefont{J.~D.} \bibnamefont{Franson}},
  \bibinfo{journal}{J.\ Mod.\ Opt.} \textbf{\bibinfo{volume}{55}},
  \bibinfo{pages}{2117} (\bibinfo{year}{2008}).

\bibitem[{\citenamefont{Reznik et~al.}(2005)\citenamefont{Reznik, Retzker, and
  Silman}}]{reznik}
\bibinfo{author}{\bibfnamefont{B.}~\bibnamefont{Reznik}},
  \bibinfo{author}{\bibfnamefont{A.}~\bibnamefont{Retzker}}, \bibnamefont{and}
  \bibinfo{author}{\bibfnamefont{J.}~\bibnamefont{Silman}},
  \bibinfo{journal}{Phys.\ Rev.\ A} \textbf{\bibinfo{volume}{71}},
  \bibinfo{pages}{042104} (\bibinfo{year}{2005}).

\bibitem[{\citenamefont{Retzker et~al.}(2005)\citenamefont{Retzker, Cirac, and
  Reznik}}]{reznikII}
\bibinfo{author}{\bibfnamefont{A.}~\bibnamefont{Retzker}},
  \bibinfo{author}{\bibfnamefont{J.~I.} \bibnamefont{Cirac}}, \bibnamefont{and}
  \bibinfo{author}{\bibfnamefont{B.}~\bibnamefont{Reznik}},
  \bibinfo{journal}{Phys.\ Rev.\ Lett.} \textbf{\bibinfo{volume}{94}},
  \bibinfo{pages}{050504} (\bibinfo{year}{2005}).

\bibitem[{\citenamefont{Le{\'o}n and Sab{\'i}n}()}]{conjuan}
\bibinfo{author}{\bibfnamefont{J.}~\bibnamefont{Le{\'o}n}} \bibnamefont{and}
  \bibinfo{author}{\bibfnamefont{C.}~\bibnamefont{Sab{\'i}n}},
   \eprint{arXiv[quant-ph]:0804.4641}.

\bibitem[{\citenamefont{Fermi}(1932)}]{fermi}
\bibinfo{author}{\bibfnamefont{E.}~\bibnamefont{Fermi}},
  \bibinfo{journal}{Rev.\ Mod.\ Phys.} \textbf{\bibinfo{volume}{4}},
  \bibinfo{pages}{87} (\bibinfo{year}{1932}).

\bibitem[{\citenamefont{Zippilli et~al.}()\citenamefont{Zippilli,
  Olivares-Renter{\'i}a, Morigi, Schuck, Rhode, and Eschner}}]{morigi}
\bibinfo{author}{\bibfnamefont{S.}~\bibnamefont{Zippilli}},
  \bibinfo{author}{\bibfnamefont{G.~A.} \bibnamefont{Olivares-Renter{\'i}a}},
  \bibinfo{author}{\bibfnamefont{G.}~\bibnamefont{Morigi}},
  \bibinfo{author}{\bibfnamefont{C.}~\bibnamefont{Schuck}},
  \bibinfo{author}{\bibfnamefont{F.}~\bibnamefont{Rhode}}, \bibnamefont{and}
  \bibinfo{author}{\bibfnamefont{J.}~\bibnamefont{Eschner}},
   \eprint{arXiv[quant-ph]:0806.1052}.

\bibitem[{\citenamefont{Mattle et~al.}(1996)\citenamefont{Mattle, Weinfurter,
  Kwiat, and Zeilinger}}]{zeilinger}
\bibinfo{author}{\bibfnamefont{K.}~\bibnamefont{Mattle}},
  \bibinfo{author}{\bibfnamefont{H.}~\bibnamefont{Weinfurter}},
  \bibinfo{author}{\bibfnamefont{P.~G.} \bibnamefont{Kwiat}}, \bibnamefont{and}
  \bibinfo{author}{\bibfnamefont{A.}~\bibnamefont{Zeilinger}},
  \bibinfo{journal}{Phys. Rev. Lett} \textbf{\bibinfo{volume}{76}},
  \bibinfo{pages}{4656} (\bibinfo{year}{1996}).

\bibitem[{\citenamefont{Cohen-Tannoudji
  et~al.}(1998)\citenamefont{Cohen-Tannoudji, Dupont-Roc, and
  Grynberg}}]{cohentannoudji}
\bibinfo{author}{\bibfnamefont{C.}~\bibnamefont{Cohen-Tannoudji}},
  \bibinfo{author}{\bibfnamefont{J.}~\bibnamefont{Dupont-Roc}},
  \bibnamefont{and} \bibinfo{author}{\bibfnamefont{G.}~\bibnamefont{Grynberg}},
  \emph{\bibinfo{title}{Atom-photon interactions}} (\bibinfo{publisher}{Wiley
  Interscience}, \bibinfo{address}{New York}, \bibinfo{year}{1998}).

\bibitem[{\citenamefont{Power and Thirunamachandran}(1997)}]{powerthiru}
\bibinfo{author}{\bibfnamefont{E.~A.} \bibnamefont{Power}} \bibnamefont{and}
  \bibinfo{author}{\bibfnamefont{T.}~\bibnamefont{Thirunamachandran}},
  \bibinfo{journal}{Phys. Rev. A} \textbf{\bibinfo{volume}{56}},
  \bibinfo{pages}{3395} (\bibinfo{year}{1997}).

\bibitem[{\citenamefont{Milonni et~al.}()\citenamefont{Milonni, James, and
  H.~Fearn}}]{milonni}
\bibinfo{author}{\bibfnamefont{P.~W.} \bibnamefont{Milonni}},
  \bibinfo{author}{\bibfnamefont{D.~F.~V.} \bibnamefont{James}},
  \bibnamefont{and} \bibinfo{author}{\bibfnamefont{H.}
  \bibnamefont{Fearn}},\bibinfo{journal}{Phys. Rev. A} \textbf{\bibinfo{volume}{52}},
  \bibinfo{pages}{1525} (\bibinfo{year}{1995}) .

\bibitem[{\citenamefont{Biswas et~al.}(1990)\citenamefont{Biswas, Compagno,
  Palma, Passante, and Persico}}]{compagnoI}
\bibinfo{author}{\bibfnamefont{A.~K.} \bibnamefont{Biswas}},
  \bibinfo{author}{\bibfnamefont{G.}~\bibnamefont{Compagno}},
  \bibinfo{author}{\bibfnamefont{G.~M.} \bibnamefont{Palma}},
  \bibinfo{author}{\bibfnamefont{R.}~\bibnamefont{Passante}}, \bibnamefont{and}
  \bibinfo{author}{\bibfnamefont{F.}~\bibnamefont{Persico}},
  \bibinfo{journal}{Phys. Rev. A} \textbf{\bibinfo{volume}{42}},
  \bibinfo{pages}{4291} (\bibinfo{year}{1990}).

\bibitem[{\citenamefont{Hill and Wootters}(1997)}]{wootters}
\bibinfo{author}{\bibfnamefont{S.}~\bibnamefont{Hill}} \bibnamefont{and}
  \bibinfo{author}{\bibfnamefont{W.~K.} \bibnamefont{Wootters}},
  \bibinfo{journal}{Phys.\ Rev.\ Lett.} \textbf{\bibinfo{volume}{78}},
  \bibinfo{pages}{5022} (\bibinfo{year}{1997}).

\end{thebibliography}
\end{document}